\documentclass[11pt, letterpaper]{article}                                                         
\usepackage[margin=1in]{geometry}

\usepackage{color}
\usepackage{silence}
\usepackage{cite}
\usepackage{algorithmic}
\usepackage{algorithm}
\usepackage{amssymb}
\usepackage{amsmath}
\usepackage{multirow}
\usepackage{multicol}   
\usepackage[justification=centering]{caption}
\usepackage{footnote}
\makesavenoteenv{tabular}
\usepackage{graphicx}

\usepackage{caption}
\usepackage{subcaption}

\usepackage{epstopdf,float,amsfonts}
\usepackage{epsfig}

\newcommand{\qed}{\hfill \ensuremath{\square}}
\begin{document}
\title{\bf Data-Driven Pole Placement in LMI Regions with Robustness Constraints}
\author{Sayak Mukherjee, Ramij R. Hossain
\thanks{S. Mukherjee is with the Optimization and Control Group, Pacific Northwest National Laboratory (PNNL), Richland, WA, USA, and R. R. Hossain is with the Department of Electrical and Computer Engineering, Iowa State University, Ames, IA, USA.
\textit{Authors have equal contributions}.
Emails: sayak.mukherjee@pnnl.gov, rhossain@iastate.edu.}}
\date{}
\maketitle
\begin{abstract}
    \noindent This paper proposes a robust learning methodology to place the closed-loop poles in desired convex regions in the complex plane. We considered the system state and input matrices to be unknown and can only use the measurements of the system trajectories. The closed-loop pole placement problem in the linear matrix inequality (LMI) regions is considered a classic robust control problem; however, that requires knowledge about the state and input matrices of the linear system. We bring in ideas from the behavioral system theory and persistency of excitation condition-based fundamental lemma to develop a data-driven counterpart that satisfies multiple closed-loop robustness specifications, such as $\mathcal{D}$-stability and mixed $H_2/H_{\infty}$ performance specifications. Our formulations lead to data-driven semi-definite programs (SDPs) that are coupled with sufficient theoretical guarantees. We validate the theoretical results with numerical simulations on a third-order dynamic system. 
\end{abstract}


\section{Introduction}

Recent research works in automatic control has been focused more on converting the classic model-based formulations to their data-driven counter-parts. Motivation of such designs are taken from the increasing complexity of practical dynamic systems with increase in their scale, making the dynamic model and parameters less accurately known along with several unmodeled non-idealities. Data-driven approaches varied in many forms with their distinct characteristics. In machine learning community, sequential decision making problems using Markov decision process (MDPs) have garnered lot of interest under the umbrella of reinforcement learning (RL) \cite{RL,Q,bertsekas,ADP1}. Many underlying concepts of RL have been translated to dynamic system viewpoint using adaptive dynamic programming in approaches such as \cite{vrabie1, jiang1, V17, V18, sayak_arxiv, mukherjee2020distributed}, considering both partial and full model-free designs. The later class of methods intends to provide theoretical guarantees using dynamic systems theory. On continuing the path of supplementing data-driven algorithms with strong mathematical backing, behavioral system theory is recently being touted as an effective alternative approach \cite{willems2005note}. The underlying idea is to represent the space of input-output trajectories of an LTI system to be spanned by single time shifted trajectory measurements. Research works such as \cite{de2019formulas, van2020data, berberich2020robust} follow such underlying framework for data-driven optimal control designs.

Along with considering optimal control designs in a data-driven way, practical systems will require sufficient robustness margins. The problem of performing robust control designs with unknown state model is currently having a lot of open questions. Approaches have been developed such as \cite{morimoto2000robust, pinto2017robust, jiang_book, sayak_acc}, that tries to augment some robustness aspects by infusing few robust considerations in the data-driven RL or optimal control setting. However, dedicated robust learning methodology is envisioned to provide much better stabilization and performance guarantees with strong underlying framework. Behavioral system theory using Willems fundamental lemma \cite{willems2005note} can provide such foundations for robust learning control designs that helps to provide an one-to-one conversion strategy from model based to data-driven approaches. In this paper we build upon that framework and consider the pole placement problem in desired convex region of the complex plane along with sufficient robustness specifications.  
\par
Classically, decades of research in the robust control domain unearthed plethora of methods that can provide sufficient system performance in presence of noise, unmodeled dynamics, uncertainty etc \cite{khargonekar1991mixed,doyle1988state,fujisaki1996linear}. The closed-loop pole placement problem in desired convex region in the complex plane has given rise to linear matrix inequalities in works such as \cite{chilali1996h}. However, these classical methods require the knowledge about system state and input matrices. With this motivation, this paper deals with the robust pole placement problem in the LMI regions under the assumption that the system state and input matrices are unknown and the designer only has access to trajectory measurements. The system is explored with persistently exciting inputs to make sure we do not violate the fundamental requirement of behavioral system theory. The model based formulations, thereafter, can be converted to data-driven formulations using the closed-loop data-driven parametrized representations. The robust control problem considers the LMI based pole placement conditions along with robust performance constraints such as mixed $H_2/H_{\infty}$ performance requirements. 

\noindent \textbf{Contribution.} The main contribution of the paper is to propose a data-driven robust control methodology that can achieve desired closed-loop pole placement in convex regions of the complex plane along with sufficient robust and optimal system performance constrained by imposing mixed $H_2/H_{\infty}$ performance metrics. We use the fundamental lemma based data-driven parametrized representation to formulate the convex formulations of the robust pole placement problem in LMI regions that can achieve very close to the model-based design characteristics. We provide theoretical guarantees on the performance of the proposed algorithm along with validation on a third order dynamic system example.  

The rest of the paper is organized as follows. Section II describes the model and problem statement considered for the paper. We recall some fundamentals for data-driven designs in Section III. The main results on the data-driven LMI region pole placement with robustness specifications are shown in Section IV. Numerical example is given in Section V, and we provide concluding remarks in Section VI. 

\noindent \textbf{Notations.} $S \succ(\succeq) 0$ denotes positive definite (semidefinite) matrix;
$H_2$ norm : The $H_2$ norm of system $G$ in time domain is given by, $||G||_2 = (\int_0^\infty [\mbox{trace}(h(t)^Th(t))]dt)^{\frac{1}{2}} $ where $h(t)$ is the impulse response; $H_\infty$ norm : The system $H_\infty$ norm is given by $||G||_\infty = \mbox{sup}_w \sigma_{\mbox{max}} (G(jw))$, where $\sigma_{\mbox{max}}$ denotes maximum singular value, and $G(jw)$ is the system transfer matrix.

\section{Problem Statement}
We consider a linear time-invariant (LTI) continuous-time dynamic system of the form:
\begin{align}\label{system}
    \dot{x} = Ax + B_1w + B_2u,\; x(0)=x_0,
\end{align}
where $x \in \mathbb{R}^{n}, u \in \mathbb{R}^{m}, w \in \mathbb{R}^{d}$ are the states and control and extraneous inputs. We, hereby, make the following assumption.\\
\textbf{Assumption 1:} The dynamic state matrix $A$ and control input matrix $B_2$ are unknown. However, the values of $n,m$ and $d$ are known.
\par
We also consider the following assumption about the availability of measurements.\\
\textbf{Assumption 2:} The measurements of states $x(t)$ and control inputs $u(t)$ are available to the designer, and the designer injects known extraneous disturbances $w(t)$ in a controlled environment to perform the control design tasks. \\
We are interested in placing the closed-loop poles in some desired regions of the complex plane. These regions are classically known as LMI regions defined as follows.\par
\noindent \textbf{Definition 1:} LMI pole placement regions - A subset $\mathcal{D}$ can be characterized as the desired pole placement regions if there exist a symmetric matrix $\alpha = [a_{ij}] \in \mathbb{R}^{n \times n}$, and a matrix $\beta = [b_{ij}] \in \mathbb{R}^{n \times n} $ such that,
\begin{align}\label{lmireg1}
    \mathcal{D}(z) = \{ z \in \mathbb{C}, \psi_\mathcal{D} < 0 \},
\end{align}
where,
\begin{align}
    \psi_\mathcal{D}(z) = \alpha + z\beta + z^*\beta^T = [a_{ij}+b_{ij}z + \beta_{ji}z^*]_{1 \leq i,j \leq n}.
\end{align}

The notation $ M = [m_{ij}]_{1 \leq i,j \leq n}$ denotes $M$ to be a $n \times n$ matrix (resp. block matrix) with generic entry (resp. block) $m_{ij}$. 
An LMI region is a subset of the complex plane that is representable by an LMI in $z$ and $z^*$, or
equivalently, an LMI in $x = Re(z)$ and $y = Im(z)$. As a
result, LMI regions are convex. Moreover, LMI regions are
symmetric with respect to the real axis. Various different types of LMI regions can be constructed by designer. Through out this paper, we consider the following as our working example.\\
\noindent \textbf{Example 1:} To consider the region in the left half of complex plane between the lines with slopes $-\frac{1}{\alpha}$ and $\frac{1}{\alpha}, \alpha > 0$ is given as,
\begin{align}\label{lmireg}
    \mathcal{D} = \{ z \in \mathcal{C}, \begin{bmatrix} z + z^* & \alpha(z^* - z)  \\
    \alpha(z - z^*) & z + z^* \end{bmatrix} < 0 \}.
\end{align}
This can be straightforwardly shown using $z = x+jy,$ giving,
\begin{align}
    \begin{bmatrix} z + z^* & \alpha(z^* - z)  \\
    \alpha(z - z^*) & z + z^* \end{bmatrix} =  \begin{bmatrix} 2x & j2\alpha y \\
    -j2\alpha y & 2x \end{bmatrix} < 0
\end{align}
Therefore using Schur complement,
\begin{align}
    x <0, \\ 2x+(j2\alpha y)\frac{1}{2x}(j2\alpha y) < 0,\\
    \text{implying,} -\frac{1}{\alpha} < \frac{y}{x} < \frac{1}{\alpha}.
\end{align}
If the sector in the left half of the complex plane is described using the inner angle $\theta$, then the LMI region expression becomes,
\begin{align}\label{lmireg2}
    \psi_{\mathcal{D}}(z;\theta) = \begin{bmatrix} \sin \theta (z + z^*) & \cos \theta (z - z^*) \\
    \cos \theta (z^* - z) & \sin \theta (z + z^*)
    \end{bmatrix} < 0.
\end{align}\qed \\
As the LMI region is convex, we can construct more complicated LMI regions by realizing convex polygons with intersection of simpler LMI regions. The focus of this paper to designer controllers without the state dynamics and using the state and input trajectory measurements. We also incorporate robust optimization objectives along with the LMI-based pole placement constraints. We consider the mixed $H_2/H_{\infty}$ optimization objective. We consider two controlled output variables along with the dynamics,
\begin{equation}\label{eqnmainsys}
\Sigma:\left\{\begin{array}{l}
\dot{x}(t)=A x(t)+B_1w(t) + B_2u(t), \\
z_{1}(t)=C_{1} x(t)+D_{11} w(t) + D_{12}u(t), \\
z_{2}(t)=C_{2} x(t)+D_{22} u(t).
\end{array}\right.
\end{equation}
$T_{wz_1}$ (respectively $T_{wz_2}$) denotes the trasnfer function from $w(t)$ to controlled output $z_1(t)$ (respectively to $z_2(t)$). We intend to learn the state-feedback control $u = Kx$ such that the poles of the underlying closed-loop dynamics $A + BK$ lie in the desired LMI region characterized by the prescribed $\psi_{\mathcal{D}}(z) < 0,$ thereby maintaining $\mathcal{D}$-stability, and to also satisfy mixed $H_2/H_{\infty}$ objectives on the regulated variables.  The problem statement is given as follows:\\
\textbf{P.} With the assumptions 1 and 2, learn the state-feedback control $u = Kx$ such that:
\begin{itemize}
    \item The prescribed $\mathcal{D}$-stability is maintained for a desired LMI region,
    \item Meet a prescribed $H_{\infty}$ robustness criterion, i.e, $||T_{wz_1}||_{\infty} < \gamma$, or minimize the $H_{\infty}$ norm assuming the robustness margin as a variable,
    \item With the desired $\mathcal{D}$-stability and $H_{\infty}$ robustness margin, minimize the $H_2$ performance $||T_{wz_2}||_2$.
\end{itemize}
\section{Data-Driven Representation Fundamentals}
\subsection{Recalling fundamental lemma}
Consider a signal $s : \mathbb{Z} \to \mathbb{R}^p$, the Hankel matrix associated with it is given as,
\begin{align}
    S_{i,L,N} = \begin{bmatrix} 
    s(i) & s(i+1) & \dots & s(i+N-1) \\
    s(i+1) & s(i+2) & \dots & s(i+N)\\
    \dots & \dots & \dots & \dots \\
    s(i+L-1) & s(i+L) & \dots & s(i+N+L-2)
    \end{bmatrix}.
\end{align}
The Hankel matrix starts with the element $s(i)$, and consists of $L$ rows and $N$ columns. With $L=1$ we denote,
\begin{align}\label{inputseq}
    S_{i,N} = \begin{bmatrix} 
    s(i) & s(i+1) & \dots & s(i+N-1) \end{bmatrix}.
\end{align}
\noindent \textbf{Definition 2 \cite{willems2005note, de2019formulas}:} The signal $s_{[0,T-1]} \in \mathbb{R}^p$ is persistently exciting of order $L$ if the corresponding Hankel matrix $S_{0,L,T-L+1}$ has full rank $pL$. Therefore, the signal
must be sufficiently extended, i.e., $T \geq (p+1)L - 1$. We recall the Willems et al.'s fundamental lemma \cite{willems2005note} for the discrete time dynamic system:
\begin{align}\label{discrete}
    x(k+1) = Ax(k) + Bu(k),\\
    y(k) = Cx(k) + Du(k), \nonumber
\end{align}
where $x \in \mathbb{R}^{n}, u \in \mathbb{R}^{m},$ and $ y \in \mathbb{R}^{p}$.

\noindent \textbf{Lemma 1 \cite{willems2005note}:} Considering the discrete-time system as given in \eqref{discrete}, when the input $u_{[0,T-1]}$ is persistently exciting of order $n+t$  then one will have,
\begin{align}
    \mbox{rank}(\begin{bmatrix} U_{[0,t,T-t+1]} \\ X_{[0,T-t+1]} \end{bmatrix}) = n + tm.
\end{align}
\noindent \textbf{Lemma 2 \cite{willems2005note}:} For the system \eqref{discrete}, if the input $u_{[0,T-1]}$ is persistently exciting of order $n+t$, then one can express any $t-$length input-output trajectory measurements of the system in the following form,
\begin{align}
    \begin{bmatrix} u_{[0,t-1]} \\ x_{[0,t-1]} \end{bmatrix}= \begin{bmatrix} U_{[0,t,T-t+1]} \\ X_{[0,t,T-t+1]} \end{bmatrix}g,
\end{align}
where $g \in \mathbb{R}^{T-t+1}$. This shows that  when $T$ is taken sufficiently large, the rank condition of Lemma 1 can be satisfied, and therefore, any input-output trajectory of the system can be represented as a linear combination of collected input/output data. This property enables us to replace a parametric description of the system with a data based counterpart. For a persistently exciting input sequence $u_{[0,T-1]}$  of order $n+1$ with $t = 1$, $T \geq (m + 1)n + m$ is necessary for the persistence of excitation condition to hold. This results in,
\begin{align}\label{rank1}
    \mbox{rank}(\begin{bmatrix} U_{[0,1,T-t+1]} \\ X_{[0,T-t+1]} \end{bmatrix}) = n + m.
\end{align}
This idea can be extended for continuous-time systems as shown in \cite{de2019formulas}. For a sampling time $\Delta > 0$, input and state-sampled trajectories $U_{[0,1,T]}$, and $X_{[0,T]}$ are stored, and the rank condition \eqref{rank1} needs to be checked. \cite{de2019formulas} constructed the continuous-time counterpart for the time-shifted states in the discrete-time using the derivative information with slight abuse of notation:
\begin{align}
    X_{1,T} = \begin{bmatrix} 
    \dot{x}(0) & \dot{x}(\Delta) & \dots & \dot{x}((T-1)\Delta) \end{bmatrix}.
\end{align}
The state-dynamic data gathered over the $T$-length window is represented as,
\begin{align}
    X_{1,T} = AX_{0,T} + BU_{0,1,T},\\
    = [B \;\;\;\; A]\begin{bmatrix}
    U_{0,1,T}\\
    X_{0,T}
    \end{bmatrix}.
\end{align}
\subsection{Data-driven Closed-loop Representation}

Following \cite{de2019formulas}, Lemma 2 can be exploited to derive a parametrization of
the closed loop system with a state-feedback law $u = Kx$. For the closed-loop system, 
\begin{align}
    \dot{x} &= Ax + Bu = (A+BK)x,
\end{align}
by the Rouché–Capelli theorem, there exists a
matrix $G \in \mathbb{R}^{T \times n}$ such that
\begin{align}\label{eqnKI}
    A+BK = [B \;\; A] \begin{bmatrix} K \\ I\end{bmatrix},\\
    \begin{bmatrix} K \\ I\end{bmatrix} = \begin{bmatrix} U_{0,1,T} \\ X_{0,T} \end{bmatrix} G. \nonumber
\end{align}
Therefore the data-driven representation becomes,
\begin{align}
    \dot{x} &= [B \;\; A] \begin{bmatrix} U_{0,1,T} \\ X_{0,T} \end{bmatrix} G x,\\
    & = X_{1,T} G x,
\end{align}
and the model-based closed-loop can now be made data-based as follows
\begin{align}\label{eqnX1G}
    A+BK = [B \;\; A]\begin{bmatrix} K \\ I\end{bmatrix} = X_{1,T}G,
\end{align}
The control now becomes $u = U_{0,1,T}Gx$. Therefore, the designer needs to learn the matrix $G$ to implement the feedback control. We now provide the main results of the paper.



\section{Data-Driven Robust Pole Placement Methodology}\label{sec4}
We first discuss the challenges in pole placement problems using the data-driven method without invoking any robust performance constraints.
Then, we introduce the formulation of data-driven pole placement problem combining $H_2/H_{\infty}$ condition followed by a comprehensive discussion. 

\indent Although the pole placement requirement can be on any convex region as given by (\ref{lmireg1}), we consider (\ref{lmireg}) as a working example throughout the methodology development in this section and numerical example in the following section. Considering the system $\dot{x} = Ax + B_2u$, in Section-II, we defined an example LMI region (\ref{lmireg}) for $\mathcal{D}$-stability. Next, we will state the LMI condition needs to be satisfied to place the closed loop poles in the prescribed region using a state-feedback control $u = Kx$.

\noindent \textbf{Lemma 3\cite{chilali1996h}:} For the system $\dot{x} = Ax + B_2u$, given an LMI region $\mathcal{D}$ defined by (\ref{lmireg}), the closed loop system $\tilde{A}=A+B_2K$ is said to be $\mathcal{D}$-stable if there exists a real symmetric matrix $X_D \succ 0$ satisfying the following condition.
\begin{align}\label{eqnD}
    \begin{bmatrix} \tilde{A} X_D + X_D \tilde{A}^T & \alpha(\tilde{A} X_D - X_D \tilde{A}^T)\\
    \alpha(X_D \tilde{A}^T  - \tilde{A}X_D) & \tilde{A} X_D + X_D \tilde{A}^T
    \end{bmatrix} \prec 0, 
\end{align} \qed\\
\noindent \textbf{Proof\cite{chilali1996h}:} The above condition can easily be obtained by replacing $z$ with $\tilde{A}X_{D}$ and $z^*$ with $X_D\tilde{A}^T$ in (\ref{lmireg}). Also, the detailed discussion of this condition can be found in \cite{chilali1996h}. This guarantees that the eigen values of $\tilde{A}=A+B_2K$ belong to the conic region $\psi_{\mathcal{D}}(z;\theta)$ on the left half of complex plane. \qed\\
\noindent Note that, the condition given in (\ref{eqnD}) is not an LMI, because if we replace $\tilde{A}$ with $A+B_2K$, we get a product term $K X_D$ of two unknown quantities $K$ and $X_D$. But, with a simple change of variable $Y=K X_D$, (\ref{eqnD})  can be converted in to an LMI condition. 

\noindent Next, we utilize the relation given in (\ref{eqnX1G}), and derive the data based condition for placing the poles of closed loop system in the region defined by the LMI condition (\ref{lmireg}).
\vspace{0.1cm}

\noindent \textbf{Theorem 1:} For the system $\dot{x} = Ax + B_2u$ with unknown state dynamics, let the input sequence $U_{0,1,T}$ is persistently exciting, i.e. the rank condition (\ref{rank1}) holds, then any matrix $Q$ satisfying (\ref{eqnDdata}), resulting in feedback gain $K= U_{0,1,T}Q(X_{0,T}Q)^{-1}$, will make the system $\mathcal{D}$-stable for the conic region defined by (\ref{lmireg}).
\begin{align}\label{eqnDdata}
    \begin{bmatrix} X_{1,T}Q + Q^T X_{1,T}^T & \alpha(X_{1,T}Q - Q^T X_{1,T}^T)\\
    \alpha(Q^T X_{1,T}^T  - X_{1,T}Q) & X_{1,T}Q + Q^T X_{1,T}^T
    \end{bmatrix} \prec 0.
\end{align} \qed \\
\noindent \textbf{Proof:} We recall {Lemma 3} giving us the condition (\ref{eqnD}), to place the closed loops of $A+B_2K$ in the desired conic region. However, we make the assumption that the system dynamics is unknown, therefore we rely upon the data driven persistency of excitation condition as defined in Section II. It can be seen from (\ref{eqnX1G}) that under a persistently exciting input, the closed loop dynamics can be represented in parameterized form derived from time evolution of system dynamics $X_{1,T}$, i.e., $A+B_2K = X_{1,T}G$. As such, we get the following inequality,
\begin{align}
    \begin{bmatrix} X_{1,T}G X_D + X_D (X_{1,T}G)^T & \alpha(X_{1,T}G X_D - X_D (X_{1,T}G)^T)\\
    \alpha(X_D (X_{1,T}G)^T  - (X_{1,T}G)X_D) & X_{1,T}G X_D + X_D (X_{1,T}G)^T
    \end{bmatrix} \prec 0
\end{align} 
\noindent To make this inequality an LMI, we consider $GX_D = Q$, resulting in (\ref{eqnDdata}). We also have $G=Q(X_D)^{-1}$. Now recalling Section II, $K$ can be computed as $K = U_{0,1,T}G = U_{0,1,T}Q(X_D)^{-1}$. Using (\ref{eqnKI}) we have $X_{0,T}G = I$, which implies, $X_D = X_{0,T}Q$. Therefore, the feedback gain turns out to be $K= U_{0,1,T}Q(X_{0,T}Q)^{-1}$. Please note that this condition completely relies upon collected data from system trajectories and the design parameter $\alpha$ which determines the LMI region (\ref{lmireg}). \qed \\
\noindent \textbf{Remark 1:} It is essential to note that the above pole placement problem is a feasibility problem. Therefore, it is apparent that there can be many feasible $X_D$ which lies in the convex set defined by (\ref{eqnD}). This gives rise to different controller gains which all satisfy the $\mathcal{D}$-stability condition. Similar characteristics will be observed during data-driven design using Theorem 1, as different exploration trajectories can result in different possible control gains, however, they all satisfy the desired closed-loop pole placement constraint \eqref{lmireg}.

 As described in \text{Remark 1}, we will now consider system (\ref{eqnmainsys}) along with desired $H_2/H_{\infty}$ constraints. Please note, we are now considering an extraneous input $w$ and transfer functions associated with the $H_2$ and $H_{\infty}$ problem represent the gain from $w$ to regulated outputs $z_2$ and $z_1$, respectively. As we consider the state matrix $A$ and input matrix $B$ are unknown, the extraneous disturbance needs to pre-specified in a controlled environment during the design of the feedback gains. Recalling Section II, the trajectory based system dynamics turns out to be  
\begin{align}\label{trajroll}
    X_{1,T} = AX_{0,T}  + B_1W_{0,T} + B_2U_{0,1,T}, \\
    X_{1,T} - B_1W_{0,T} = AX_{0,T} + B_2U_{0,1,T}.
\end{align}
\noindent Therefore, the closed loop parameterized representation modifies to (\ref{X1_tildeG}). Note, like $X_{0,T}$, $W_{0,T}$ can also be defined using (\ref{inputseq}). The closed loop parameterization with extraneous input becomes
\begin{align}\label{X1_tildeG}
    A+B_2K = [B_2 \;\; A]\begin{bmatrix} K \\ I\end{bmatrix} =  (X_{1,T} - B_1W_{0,T})G = \Tilde{X}_{1,T}G.
\end{align}
\noindent We now state the following theorem to solve problem \textbf{P}. 
\vspace{0.1cm}

\noindent \textbf{Theorem 2:} For the system (\ref{eqnmainsys}), to place the closed loop poles in the desired LMI region (\ref{lmireg}) along with sufficient mixed $H_2/H_{\infty}$ the following set of data driven LMIs needs to be solved, where $ C_2 = \begin{bmatrix} Q_{x}^{\frac{1}{2}} \\0 \end{bmatrix}$, $ D_{22} = \begin{bmatrix} 0 \\R^{\frac{1}{2}} \end{bmatrix}$ and $Q_x \succeq 0$, $R \succ 0$ are the designable state and input penalty factors.

\begin{equation}\label{mainddobj}
\min _{Q, S ,\gamma} \operatorname{trace}\left(Q_{x} X_{0,T} Q\right)+\operatorname{trace}(S) + \gamma
\end{equation}
subject to
\begin{align}
\begin{bmatrix}\label{hinfdata}
 \Tilde{X}_{1,T}Q+Q^T \Tilde{X}_{1,T}^T & B_1 & X_{0,T}QC_1^T+Q^TU_{0,1,T}^TD_{12} \\
B_1^{T} & -\gamma I &D_{11}^{T} \\
C_1X_{0,T}Q+D_{12}U_{0,1,T}Q & D_{11} & -\gamma I
\end{bmatrix}\prec0,
\end{align}
\normalsize
\begin{align}\label{ddpole}
    \begin{bmatrix} X_{1,T}Q + Q^T X_{1,T}^T & \alpha(X_{1,T}Q - Q^T X_{1,T}^T)\\
    \alpha(Q^T X_{1,T}^T  - X_{1,T}Q) & X_{1,T}Q + Q^T X_{1,T}^T
    \end{bmatrix} \prec 0,
\end{align}
\begin{align}\label{h2objschur}
    \begin{bmatrix} S & R^{1 / 2}U_{0,1,T}Q\\
Q^TU_{0,1,T}^TR^{1/2} & X_{0,T}Q
    \end{bmatrix}\succeq 0.
\end{align}
\normalsize

\noindent \textbf{Proof:} We start with considering the $H_2$ performance objective. Please note in (\ref{eqnmainsys}), $H_2$ performance objective is to minimize $||T_{wz_2}||_2$. Following \cite{feron1992numerical,boyd1994linear}, the model-based optimization problem for the $H_2$ performance is given as follows:
\begin{equation}\label{h2obj}
\min _{K, X_2} \operatorname{trace}\left(Q_{x} X_2\right)+\operatorname{trace}(R^{\frac{1}{2}} K X_{2} K^{T} R^{\frac{1}{2}})
\end{equation}
subject to
$$
\left\{\begin{array}{l}
(A+B_2 K) X_2 + X_2(A+B_2 K)^{T} + B_1 B_1^T \prec 0, \\
X_2 = X_2^T > 0.
\end{array}\right.
$$

\noindent We now consider the $H_{\infty}$ performance objective which intends to minimize $||T_{wz_1}||_{\infty}$, and the use of KYP lemma (Bounded real lemma) results into the following optimization problem, \cite{scherer2000linear,gahinet1994linear,iwasaki1994all}
\begin{equation}
\min _{\gamma,X_{\infty}} \gamma
\end{equation}
subject to
\begin{align}
\begin{bmatrix}
\tilde{A} X_{\infty}+X_{\infty} \tilde{A}^{T} & B_1 & X_{\infty} (C_1 + D_{12}K)^{T} \\
B_1^{T} & -\gamma I & D_{11}^{T} \\
(C_1 + D_{12}K) X_{\infty} & D_{11} & -\gamma I 
\end{bmatrix} \prec 0, \\\text{where,}\; \tilde{A} = A+B_2K,\nonumber \normalsize
\end{align}
\begin{align}
    X_{\infty} \succ 0.
\end{align}
\noindent Recalling Lemma 3, the pole placement condition needs to satisfy the following inequality,
\begin{align}
    \begin{bmatrix} \tilde{A} X_D + X_D \tilde{A}^T & \alpha(\tilde{A} X_D - X_D \tilde{A}^T)\\
    \alpha(X_D \tilde{A}^T  - \tilde{A}X_D) & \tilde{A} X_D + X_D \tilde{A}^T
    \end{bmatrix} \prec 0, \nonumber \\\text{where,}\; \tilde{A} = A+B_2K \nonumber.
\end{align} 
\noindent Next, we convert the above conditions in an optimization problem by seeking a common solution of $X$, where $X = X_2 = X_{\infty} = X_{D} \succ 0$. Using a single Lyapunov matrix X that enforces multiple constraints has been studied in \cite{chilali1996h,scherer2000linear}. As we are considering the mixed $H_2/H_{\infty}$ objective the stabilization constraint provided by the $H_{\infty}$ problem will serve as a conservative unifying condition for both $H_2$ and $H_{\infty}$ problem. Therefore, the model-based solution of problem \textbf{P}, can be written as,
\begin{equation}\label{modobj}
\min _{K, X, S, \gamma} \operatorname{trace}\left(Q_{x} X\right)+\operatorname{trace}(S) + \gamma
\end{equation}
subject to
\begin{align}\label{hinf}
\begin{bmatrix}
\tilde{A} X+X \tilde{A}^{T} & B_1 & X (C_1 + D_{12}K)^{T}\\
B_1^{T} & -\gamma I & D_{11}^{T} \\
(C_1 + D_{11}K) X & D_{11} & -\gamma I
\end{bmatrix} \prec 0, \\
\text{where,}\; \tilde{A} = A+B_2K,\nonumber
\end{align}
\begin{align}\label{polemd}
    \begin{bmatrix} \tilde{A} X + X \tilde{A}^T & \alpha(\tilde{A} X - X \tilde{A}^T)\\
    \alpha(X \tilde{A}^T  - \tilde{A}X) & \tilde{A} X + X \tilde{A}^T
    \end{bmatrix} \prec 0, \nonumber \\\text{where,}\; \tilde{A} = A+B_2K, 
\end{align}
\begin{align}
    X = X^T > 0,\\
S-R^{1 / 2} K X K^{T} R^{1 / 2} \succeq 0.\label{h2objschur2}
\end{align}
\noindent We represented the second term in the objective of (\ref{h2obj}) by the second term of (\ref{modobj}) and the corresponding inequality (\ref{h2objschur2}). To this end, we now move into converting these model-based expressions to their data-driven counter part. Considering (\ref{hinf}) and using (\ref{X1_tildeG}), we can have, 
\begin{equation}
\begin{bmatrix}
(\Tilde{X}_{1,T}G) X+X (\Tilde{X}_{1,T}G^{T} & B_1 & XC_1 + XK^{T} D_{12}^T\\
B_1^{T} & -\gamma I & D_{11}^{T} \\
C_1X + D_{12}KX & D_{11} & -\gamma I
\end{bmatrix} \prec 0.
\end{equation}

\normalsize
\noindent We now substitute, $GX = Q$ and $K = U_{0,1,T}G$, this results in $XK^TD_{12}^T = Q^TU_{0,1,T}^TD_{12}^T$, giving us (\ref{hinfdata}). The substitution of $K = U_{0,1,T}G$ also converts (\ref{h2objschur2}) into (\ref{h2schur3}).
\begin{align}\label{h2schur3}
    S-R^{1 / 2} U_{0,1,T}G X G^{T}U_{0,1,T}^T R^{1 / 2} \succeq 0.
\end{align}
\noindent Next, replace $GX = Q$ and $G^T = X^{-T}Q^T =(X_{0,T}Q)^{-T}Q^{T}$. Note, from (\ref{eqnKI}) we can write $X_{0,T}G = I$, therefore pre-multiplying $GX = Q$ with $X_{0,T}$ results in $X = X_{0,T}Q$. Now, we have $S-R^{1 / 2} U_{0,1,T}Q (X_{0,T}Q)^{-T}Q^{T}U_{0,1,T}^T R^{1 / 2} \succeq 0$, and after applying Schur complement, we get (\ref{h2objschur}). The first term of objective(\ref{modobj}) uses the relation $X = X_{0,T}Q$ and converts into first term of (\ref{mainddobj}). Finally, we supplement these LMI conditions with the data-driven $\mathcal{D}$-stability condition provided in Theorem 1, and this completes the proof of {Theorem 2}. \qed \\
\noindent \textbf{Remark 2} Please note that Theorem 2 considers the $H_{\infty}$ performance margin $\gamma$ as an optimization variable. However, in many scenarios the designer may be interested in a pre-specified robustness performance gain $(\bar{\gamma}) \geq \gamma_{min}$, where $\gamma_{min}$ is the solution of the problem in Theorem 2. To use a pre-specified $(\bar{\gamma})$, the objective function in Theorem 2 modifies into $\operatorname{trace}\left(Q_{x} X_{0,T} Q\right)+\operatorname{trace}(S)$ along with the $\gamma$ in (\ref{hinfdata}) will be replaced by $(\bar{\gamma})$.

\section{Numerical Example}
\begin{figure}[htbp]
  \centering
    \includegraphics[width=0.40\textwidth]{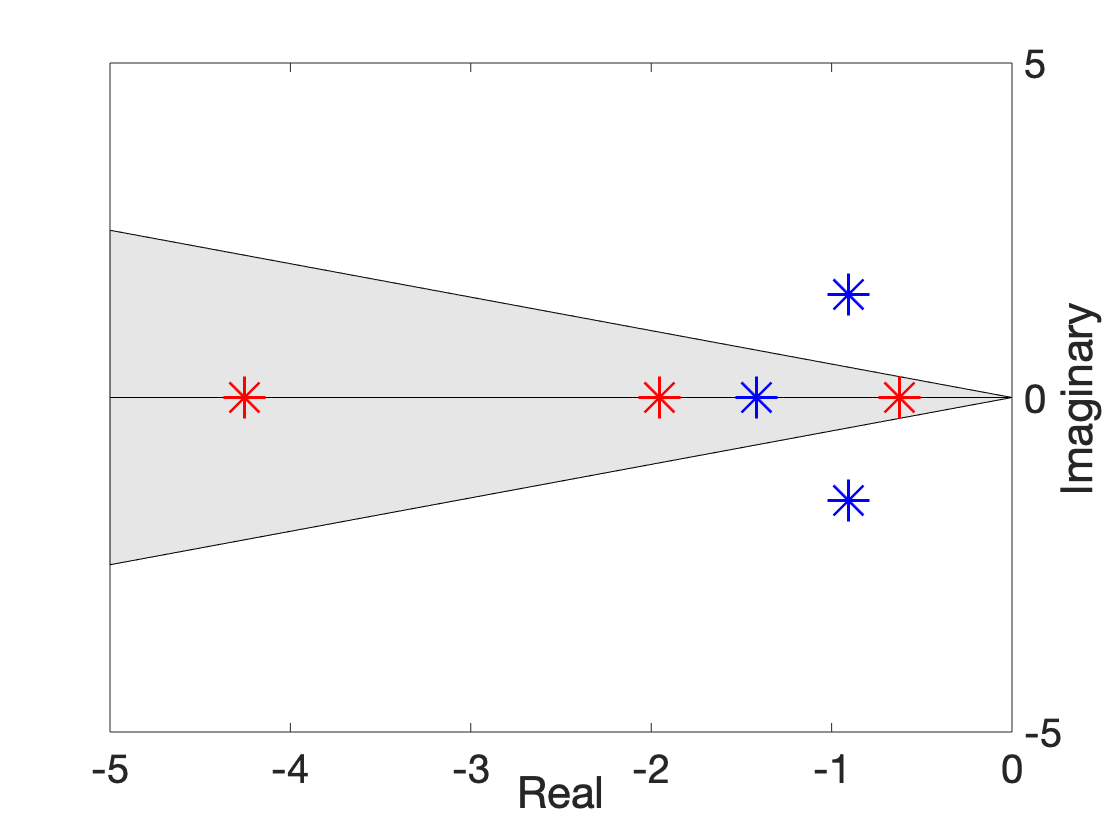}
  \caption{ Data-Driven Pole Location with (in red) and without LMI Constraints (in blue)}
  \label{f1}
 \end{figure}

In this section, we present an illustrative example of a pole placement problem with robust performance criteria using the data-driven method discussed in Section \ref{sec4}. We compare the obtained results using the data-driven approach with its model-based counterpart. We define the system given in (\ref{eqnmainsys}) with the following state and input matrices, taken from \cite{berberich2020robust}:
\[
A = \begin{bmatrix}
    -0.5 & 1.4 & 0.4 \\ 
    -0.9 & 0.3 & -1.5 \\ 
    1.1 & 1 & -0.4 
\end{bmatrix};
 B_2 =
\begin{bmatrix}
    0.1 & -0.3 \\ 
    -0.1 & -0.7 \\ 
    0.7 & -1      
\end{bmatrix} 
\].
\noindent Two of the eigenvalues of $A$ are located on the right side of the $j\omega$ axis resulting in an unstable open-loop system. Note while designing the data-driven control, it is assumed that $A$ and $B_2$ are \textit{unknown}. We choose the system matrix $B_1$ for extraneous inputs $w$ and other performance matrices $C_1, D_{11}, D_{12}, C_{2}$ and $D_{22}$ as follows: 
\[
B_1 = \begin{bmatrix}
    1 & 0 & 0 \\ 
    0 & 1 & 0 \\ 
    0 & 0 & 1 
\end{bmatrix};
C_1 = \begin{bmatrix}
    1 & 0 & 0 \\ 
    0 & 1 & 0 \\ 
    0 & 0 & 1 
\end{bmatrix};
D_{11} = \begin{bmatrix}
    1 & 1 & 1 \\ 
    1 & 1 & 1 \\ 
    1 & 1 & 1 
\end{bmatrix};
D_{12} = \begin{bmatrix}
    1 & 1 & 1 \\ 
    1 & 1 & 1 \\ 
\end{bmatrix}
\],
\[
C_{2} = \begin{bmatrix} Q_{x}^{\frac{1}{2}} 
\\\mathbf{0} \end{bmatrix}, \;\; \text{where}\;\; Q_x = \begin{bmatrix}
    1 & 0 & 0 \\ 
    0 & 1 & 0 \\ 
    0 & 0 & 1 
\end{bmatrix};
D_{22} = \begin{bmatrix} \mathbf{0} \\R^{\frac{1}{2}} \end{bmatrix}, \;\; \text{where}\;\;R = \begin{bmatrix}
    1 & 0\\ 
    0 & 1\\ 
\end{bmatrix}
\].
\noindent We choose a random initial condition $x_0$ and a random input sequence $u \in \mathbb{R}^2$ with the magnitudes of both channels constrained at $[-0.5,0.5]$. Here, the interesting part is in deciding the length of the input sequence $u$. For this example $n = 3$, $m = 2$, therefore to satisfy the rank condition given in (\ref{rank1}) as a requirement of the persistently excitation, the length of the input sequence $T$ is selected as $15$, as we require $T \geq (m + 1)n + m$. Next, to generate the trajectory rollout as defined in (\ref{trajroll}), we choose $w$ randomly from a norm ball defined by $\lvert \lvert w \rvert\rvert_2 \leq 0.05$. Trapezoidal approximations are used to generate the rollouts using continuous time dynamics. Now, for a given $\alpha = 2$, the solution of problem \textbf{P} using Theorem 2 is as follows, 
\[
K_{\text{data}} = \begin{bmatrix}
   -3.627984  &  1.257298  & -3.803731 \\
    1.433545  &  0.837496  &  2.065323\end{bmatrix}, \\ \gamma_{\text{min}} = 4.832\]
\noindent The poles of the closed-loop system are -4.2545, -1.9539, and -0.6244. These poles (marked as red* in Fig.~\ref{f1}) are located on the negative real axis, which includes the conic region defined by $\alpha = 2$ (shaded area in Fig.~\ref{f1}). To verify our designed data-driven controller gain and the location of closed-loop poles, we run the same experiments with the model-based equations given in (\ref{modobj}) to (\ref{h2objschur2}). The computed model-based gain is shown below:
\[K_{\text{mod}} = \begin{bmatrix}
  -3.627994  & 1.257302 & -3.803737 \\
   1.433540  & 0.837497  & 2.065325\end{bmatrix}, \\ \gamma_{\text{min}} = 4.832\]
\noindent The computed gain $K_{\text{data}}$ and $K_{\text{mod}}$ are identical with a difference $\lvert \lvert K_{\text{data}} - K_{\text{mod}}\rvert \rvert \approx 10^{-4}$, which validates the accuracy of our proposed data-driven design.
 
We do an ablation study to further check the robustness of the data driven method. We removed the pole placement constraints from the conditions given in Theorem 2. This simply converts the problem into a mixed $H_2/H_{\infty}$ problem. Solving the LMIs (\ref{mainddobj}) to (\ref{h2objschur}), except (\ref{ddpole}), we obtained,
\[
\bar{K}_{\text{data}} = \begin{bmatrix}
   -3.627984  &  1.257298  & -3.803731 \\
    1.433545  &  0.837496  &  2.065323\end{bmatrix},
    \]
    \[\text{Poles:} -0.9062 + j1.533,-0.9062 - j1.533, -1.4182\]
\noindent Fig.~\ref{f1} clearly indicates that these poles (marked as blue*) are located outside the LMI region (shaded area). Like previous experiments, the same results can be found in case of model-based optimization using (\ref{modobj}) to (\ref{h2objschur2}) eliminating (\ref{polemd}). 
\section{Conclusion}
In this paper, we have presented a comprehensive data-driven methodology that satisfies multiple constraints comprising of $\mathcal{D}$-stability, and mixed $H_2/H_{\infty}$ performance guarantees. We have shown that the data-based parametrized representation of closed-loop dynamics originating from the behavioral system theory can provide a fundamental framework to solve such classic problems with unknown state and input matrices. The solutions from the proposed semi-definite programs match closely with the classical model-based solutions, which has been proven rigorously and validated numerically. Future research work will consider unknown dynamic systems coupled with structured and parametric uncertainty, and develop multiple data-driven robust control designs.
\bibliographystyle{IEEEtran}
\bibliography{ref}
\end{document}